\documentclass{article}
\usepackage{graphicx}
\usepackage{epsfig}
\usepackage{url}
\usepackage{psfrag}
\usepackage{times}
\usepackage{amsfonts,amssymb,amsbsy}
\usepackage{amsmath}
\usepackage{cite}
\usepackage{latexsym}
\usepackage{amsfonts,amsbsy}
\usepackage{xcolor} 

\begin{document}

\title{Generalization of Perfect Electromagnetic Conductor (PEMC) Boundary}
\author{Ismo V. Lindell and Ari~Sihvola}%
        \date{School of Electrical Engineering,\\ Aalto University, Espoo, Finland\\ 
{\tt ismo.lindell@aalto.fi}\\\vspace{-1pt}{\tt ari.sihvola@aalto.fi}}
\pagestyle{myheadings}



\def\e{\begin{equation}} 
\def\f{\end{equation}} 
\def\ea{\begin{eqnarray}} 
\def\fa{\end{eqnarray}} 

\def\##1{{\mbox{\textbf{#1}}}}
\def\%#1{{\mbox{\boldmath $#1$}}}
\def\=#1{{\overline{\overline{\mathsf #1}}}}
\def\RR{\mbox{\boldmath $\R$}}
\def\nn#1{{\sf #1}}
\def\SE{{\mathbb E}}
\def\SF{{\mathbb F}}

\def\*{^{\displaystyle*}}
\def\xx{\displaystyle{{}^\times}\llap{${}_\times$}}
\def\.{\cdot}
\def\x{\times}
\def\oo{\infty}

\def\D{\nabla}
\def\d{\partial}

\def\ra{\rightarrow}
\def\lra{\leftrightarrow}
\def\Ra{\Rightarrow}
\def\le{\left(}
\def\ri{\right)}
\def\l#1{\label{eq:#1}}
\def\r#1{(\ref{eq:#1})}
\def\am{\left(\begin{array}{c}}
\def\amm{\left(\begin{array}{cc}}
\def\ammm{\left(\begin{array}{ccc}}
\def\ammmm{\left(\begin{array}{cccc}}
\def\a{\end{array}\right)}

\def\I{\int\limits}
\def\OI{\oint\limits}

\def\A{\alpha}
\def\B{\beta}
\def\de{\delta}
\def\De{\Delta}
\def\E{\epsilon}
\def\g{\gamma}
\def\G{\Gamma}
\def\h{\eta}
\def\K{\kappa}
\def\la{\lambda}
\def\La{\Lambda}
\def\M{\mu}
\def\o{\omega}
\def\Om{\Omega}
\def\R{\rho}
\def\s{\sigma}
\def\t{\tau}
\def\z{\zeta}
\def\X{\chi}
\def\TH{\theta}
\def\Th{\Theta}
\def\VF{\varphi}
\def\VR{\varrho}
\def\VT{\vartheta}
\def\ve{\%\varepsilon}

\def\tr{{\rm tr }}
\def\spm{{\rm spm}}
\def\det{{\rm det}}
\def\Det{{\rm Det}}
\def\sgn{{\rm sgn}}
\def\bi{\bibitem}

\def\W{\wedge}
\def\WW{\displaystyle{{}^\wedge}\llap{${}_\wedge$}}
\def\Adj{{\rm Adj\mit}}
\def\ua{\uparrow}
\def\da{\downarrow}
\def\uda{\updownarrow}

\def\J{\rfloor}
\def\L{\lfloor}
\def\JJ{\rfloor\rfloor}
\def\LL{\lfloor\lfloor}

\maketitle\

\begin{abstract}
Certain classes of electromagnetic boundaries satisfying linear and local boundary conditions can be defined in terms of the dispersion equation of waves matched to the boundary. A single plane wave is matched to the boundary when it satisfies the boundary conditions identically. The wave vector of a matched wave is a solution of a dispersion equation characteristic to the boundary. The equation is of the second order, in general. Conditions for the boundary are studied under which the dispersion equation is reduced to one of the first order or to an identity, whence it is satisfied for any wave vector of the plane wave. It is shown that, boundaries associated to a dispersion equation of the first order, form a natural generalization of the class of perfect electromagnetic conductor (PEMC) boundaries. As a consequence, the novel class is labeled as that of generalized perfect electromagnetic conductor (GPEMC) boundaries. As another case, boundaries for which there is no dispersion equation (NDE) for the matched wave (because it is an identity) are labeled as NDE boundaries. They are shown to be special cases of GPEMC boundaries. Reflection of the general plane wave from the GPEMC boundary is considered and an analytic expression for the reflection dyadic is found. Some numerical examples on its application are presented for visualization.
\end{abstract}



\section{Introduction}

Boundary conditions are known to play an essential role in defining electromagnetic field problems. It has been recently pointed out that the most general form of linear and local conditions for electromagnetic boundaries, valid at a surface with normal unit vector $\#n$, can be expressed in terms of four dimensionless vectors $\#a_1 \cdots \#b_2$ as \cite{PIER,AP17,GBC}
\ea \#a_1\.\#E + \#b_1\.\h_o\#H &=& 0, \l{GBC1}\\
\#a_2\.\#E + \#b_2\.\h_o\#H &=& 0. \l{GBC2}\fa
Here we assume for simplicity that the boundary surface is planar, i.e., that $\#n$ is a constant real unit vector. Also, the medium above the boundary is assumed to be isotropic as defined by the parameters $\M_o,\E_o$, with $\h_o=\sqrt{\M_o/\E_o}$ and $k_o=\o\sqrt{\M_o\E_o}$. The boundary defined by the general boundary conditions (GBC) \r{GBC1}, \r{GBC2} has been labeled as the GBC boundary \cite{AP17}.

Conditions \r{GBC1} and \r{GBC2} include a number of well-known boundaries as special cases. Denoting vectors tangential to the boundary surface by the subscript $()_t$, a few of them can be listed as follows \cite{GBC}.

\begin{itemize}
\item The impedance boundary, defined by four vectors tangential to the boundary surface, $\#a_i=\#a_{it},\  \#b_i=\#b_{it}$, $i=1,2$, or, more compactly as  \cite{Methods,Hoppe}
\e \#E_t=\=Z_t\.(\#n\x\#H), \l{imp}\f
with \cite{GBC}
 \e \=Z_t = \frac{-\h_o}{\#n\.\#a_{1t}\x\#a_{2t}}\#n\#n\xx(\#a_{1t}\#b_{2t}- \#a_{2t}\#b_{1t}). \f
\item The soft-and-hard (SH) boundary \cite{Kildal,SH}, defined by $\#a_1=\#b_2=\#a_t$, $\#b_1=\#a_2=0$, or,
\e \#a_t\.\#E=\#a_t\.\#H=0. \f
\item The DB boundary \cite{Rumsey,DB,259}, defined by $\#a_1=\#b_2=\#n$,\ \  $\#b_1=\#a_2=0$, or,
\e \#n\.\#E=\#n\.\#H=0. \f
\item The soft-and-hard/DB (SHDB) boundary \cite{SHDB}, a generalization to the SH and DB boundaries, defined by $\#a_1=\#a_t,\ \#b_1= \A\#n$ and $\#a_2=\A\#n,\ \#b_2=-\#a_t$, or
\ea \#a_t\.\#E + \A\#n\.\h_o\#H &=& 0, \\
\A\#n\.\#E - \#a_t\.\h_o\#H &=& 0. \fa
\item The perfect electromagnetic conductor (PEMC) \cite{PEMC,253,264}, defined by $\#b_{1,2}=(1/M\h_o)\#a_{1,2}$ and $\#a_1\x\#a_2=\#n$, or  
\e \#n\x(\#H+M\#E)=0. \l{PEMC} \f
It has the special cases of PMC $(M=0)$ and PEC $(|M|\ra\oo)$ boundaries. Also, \r{PEMC} is a special case of the impedance boundary condition \r{imp} with $\=Z_t=(1/M)\#n\x\=I$.
\end{itemize}

In \cite{GBC}, additional special cases of \r{GBC1} and \r{GBC2}, have been discussed. In the past, many of the boundaries have been given realizations in terms of physical structures \cite{Caloz13,Elmaghrabi,235,Wallen,257,258,Zaluski11,274,279,Zaluski14}. Also, many of the boundaries have recently found applications and generalizations \cite{Kildal09,234,241,244,245,251,Sihvola06,250,Sihvola07,260,280,285,Nosrati}.

\section{Matched Waves}

A plane wave is called matched to a boundary when it satisfies the boundary conditions identically. Thus, there is no reflected wave when the incident wave is matched to the boundary. Surface waves associated to impedance boundaries serve as examples of matched waves.

Conditions for matched waves at the GBC boundary are obtained by writing the relation between fields of a plane wave,
\e \#k\x\#E = k_o\h_o\#H, \f
and requiring that the plane wave field $\#E$ satisfy the boundary conditions \r{GBC1} and \r{GBC2} as
\ea (k_o\#a_1 + \#b_1\x\#k)\.\#E &=& 0, \l{E1}\\
(k_o\#a_2 + \#b_2\x\#k)\.\#E &=& 0, \l{E2}\\
\#k\.\#E &=& 0. \l{E3}\fa
Condition \r{E3} is satisfied by any plane wave.

\subsection{Dispersion Equation}

For a solution $\#E\not=0$, the three vectors in \r{E1}, \r{E2} and \r{E3} must be coplanar, i.e., they must satisfy
\e (k_o\#a_1 + \#b_1\x\#k)\x(k_o\#a_2 + \#b_2\x\#k)\.\#k=0. \l{D} \f
\r{D} restricts the choice of the wave vector $\#k$ and it is called the dispersion equation for a matched wave \cite{AP17}. After finding the solution $\#k$ of \r{D}, the field of the matched wave can be expressed in the form
\e \#E = E\#k\x(k_o\#a_1 + \#b_1\x\#k), \l{dispE}\f
in terms of some scalar factor $E$.

In spite of its cubic form, the dispersion equation \r{D} can be expanded in a form which is actually quadratic in $\#k$ \cite{GBC},
$$ (\#a_1\#b_2-\#b_1\#a_2):\#k\#k + k_o(\#a_1\x\#a_2  +\#b_1\x\#b_2)\.\#k +$$
\e +k_o^2(\#a_2\.\#b_1-\#a_1\.\#b_2)=0. \l{disp}\f
Another form for the dispersion equation is \cite{GBC}
$$ (\#a_1\x\#k)\.(\#b_2\x\#k) - (\#b_1\x\#k)\.(\#a_2\x\#k) +$$
\e -k_o(\#a_1\x\#a_2+ \#b_1\x\#b_2)\.\#k=0. \l{disp1}\f

Because the wave vector in the simple-isotropic medium is known to satisfy $\#k\.\#k=k_o^2$, we can write
\e \#k = k_o\#u,\ \ \ \ \#u\.\#u=1,\f
whence \r{disp} and \r{disp1} actually restrict the choice of the unit vector $\#u$. In the general case, $\#u$ is a complex vector, corresponding to exponential spatial dependence of the electric and magnetic fields.

\subsection{Special Cases}

The dispersion equation \r{disp} depends on the four vectors defining the boundary. For the special cases listed in the Introduction, the dispersion equation takes the following simplified forms.
\begin{itemize}
\item For the impedance boundary, the dispersion equation becomes
$$ (\#a_{1t}\#b_{2t}-\#b_{1t}\#a_{2t}):\#k_t\#k_t + k_o(\#a_{1t}\x\#a_{2t}  + \#b_{1t}\x\#b_{2t})\.\#n k_n +$$
\e + k_o^2(\#a_{2t}\.\#b_{1t}-\#a_{1t}\.\#b_{2t})=0, \f
or \cite{GBC},
\e k_ok_n(\h_o^2+ \det_t\=Z_t) + \h_o(\=Z_t:\#k_t\#k_t + k_n^2\tr\=Z_t)=0, \f
where $\det_t\=Z_t$ denotes the 2D determinant. The equation can be solved for $k_n/k_o$ in terms of given vectors $\#k_t/k_o$, thus defining the locus of the $\#k$ vector of possible matched waves.
\item For the SH boundary, with $\#a_1=\#b_2=\#a_t$, $\#a_2=\#b_1=0$, the dispersion equation \r{disp1} becomes
\e (\#a_t\x\#k)\.(\#a_t\x\#k)=  0. \f
If $\#a_t$ is a real unit vector, matched waves propagate along the boundary as $\#k=\pm \#a_t k_o$.
\item For the DB boundary, the dispersion equation is reduced to
\e (\#n\.\#k)^2 = k_o^2(\#n\.\#u)^2 = k_o^2. \f
Real solutions are $\#k=\pm \#n k_o$, which correspond to propagation normal to the DB boundary.
\item For the PEC boundary, the dispersion equation becomes
\e (\#a_{1t}\x\#a_{2t})\.\#k=0, \f
which is satisfied for any $\#k$ satisfying $\#n\.\#k=0$, i.e., for lateral waves propagating along the boundary surface.
\end{itemize}

The form \r{disp} of the dispersion equation suggests defining three classes of boundaries in terms of the order of the dispersion equation:
\begin{enumerate}
\item \r{disp} is of the second order in $\#k$
\item \r{disp} is of the first order in $\#k$
\item \r{disp} is an identity, satisfied by any $\#k=k_o\#u$.
\end{enumerate}
Actually, each class contains those below as special cases. Let us study restrictions to the boundary vectors $\#a_1\cdots \#b_2$ corresponding to the cases 2 and 3.

\section{First-Order Dispersion Equation}

For the dispersion equation \r{disp} to be of the first order, the quadratic term must vanish. Denoting
\e \=A=\#a_1\#b_2-\#b_1\#a_2, \f
the dyadic $\=A$ must satisfy 
\e \=A:\#k\#k =\frac{1}{2}(\=A+\=A{}^T):\#k\#k  =0,  \l{Akk}\f
for any (possibly complex) vector $\#k=k_o\#u$. Actually, \r{Akk} can be required to be valid for any vector $\#k$ without restriction. Choosing $\#k=\#k_1+\#k_2$, \r{Akk} yields $(\=A+\=A{}^T):\#k_1\#k_2=0$ for any two vectors $\#k_1,\#k_2$, which requires $\=A+\=A{}^T=0$. Thus, for the dispersion equation to be of the first order, the dyadic $\=A$ must be antisymmetric, whence the four vectors must satisfy the condition
\e \#a_1\#b_2+\#b_2 \#a_1=\#b_1\#a_2+\#a_2\#b_1.\l{cond2}\f
Because this implies
\e \#a_1\.\#b_2 = \#b_1\.\#a_2, \l{cond3}\f
the first and last terms of \r{disp} vanish simultaneously. The resulting first-order dispersion equation then becomes
\e (\#a_1\x\#a_2+ \#b_1\x\#b_2)\.\#k=0, \l{a1a2b1b2k}\f
where the four vectors are restricted by the condition \r{cond2}.

Dot-multiplying both sides of \r{cond2} by a vector from the left or from the right, leads to the conclusion that the two vector pairs $\#a_1,\#b_2$ and $\#a_2,\#b_1$ must be coplanar. Thus, there must exist relations of the form
\ea \#a_2 &=& A_2\#a_1 + B_2\#b_2, \\
\#b_1 &=& A_1\#a_1+ B_1\#b_2. \fa
Inserting these, the condition \r{cond2} becomes
$$ 2A_1A_2\#a_1\#a_1 + 2B_1B_2\#b_2\#b_2 + $$
\e +(A_1B_2+ A_2B_1-1)(\#a_1\#b_2+ \#b_2\#a_1)=0. \l{condi}\f
Assuming $\#a_1$ and $\#b_2$ linearly independent (otherwise all four vectors are multiples of the same vector, whence \r{a1a2b1b2k} is identically satisfied), \r{condi}  leads to the relations
\e A_1A_2=0,\ \ \ \ B_1B_2=0,\ \ \ \ A_1B_2+ A_2B_1=1, \f
which have two possible solutions,
\ea  A_1=B_2=0,\ \ &&\ \ A_2=1/B_1,\\
\#a_2=A_2\#a_1,\ \ && \ \ \#b_1=\#b_2/A_2, \l{1}\fa
and
\ea  A_2=B_1=0,\ \ &&\ \ A_1=1/B_2,\\
 \#a_2=B_2\#b_2,\ \ && \ \ \#b_1=\#a_1/B_2. \l{2}\fa

Corresponding to the case \r{1}, the boundary conditions \r{GBC1}, \r{GBC2} take the respective form
\ea \#a_1\.A_2\#E + \#b_2\.\h_o\#H &=& 0, \l{GBC1'}\\
\#a_1\.A_2\#E + \#b_2\.\h_o\#H &=& 0, \l{GBC2'}\fa
which are the same condition. Since they do not uniquely define a boundary, we can ignore this case. 

For the case \r{2}, the boundary conditions become
\ea \#a_1\.\le B_2\#E + \h_o\#H\ri &=& 0, \l{GBC1''}\\
\#a_2\.\le B_2\#E + \h_o\#H\ri &=& 0. \l{GBC2''}\fa
To have two distinct conditions, we must assume
\e \#m = \#a_1\x\#a_2\not=0. \l{m}\f

\section{Generalized PEMC Boundary}

The boundary conditions \r{GBC1''} and \r{GBC2''} can be written compactly as
\e \#m\x(\#H + M\#E)=0,\ \ \ \ M=B_2/\h_o.  \l{GPEMC} \f
Because, for $\#m=\#n$, \r{GPEMC} equals the PEMC boundary condition \r{PEMC}, we can call the boundary defined by \r{GPEMC} by the name generalized PEMC (GPEMC) boundary. Here we must note that $\#m$ need not be a real vector.

The dispersion equation \r{a1a2b1b2k} restricting the $\#k$ vectors for waves matched to the GPEMC boundary is reduced to
\e (B_2+ \frac{1}{B_2})\#m\.\#k=0. \l{condk}\f
The case $B_2^2=-1$ will be considered in the following Section. In the more general case, the linear dispersion equation must be of the simple form
\e \#m\.\#k=0,\l{mk0} \f
whence the $\#k$ vector can be expressed as
\e \#k = k_1\#a_1+ k_2\#a_2. \l{ka1a2}\f
Because of the limitation $\#k\.\#k=k_o^2$, there is one free (complex) parameter left in the representation \r{ka1a2}. For a real vector $\#m$ the real and imaginary parts of the $\#k$ vectors of possible matched waves lie in the plane orthogonal to $\#m$. For the special case of the PEMC boundary  with $\#m=\#n$, any lateral plane wave satisfying $\#n\.\#k=0$, is known to be a matched wave \cite{GBC}.
  
To interpret the boundary defined by \r{GBC1''} and \r{GBC2''}, let us consider the duality transformation of fields defined by \r{dual} and \r{ADBC} in the Appendix, known to keep the isotropic medium invariant. Because the boundary conditions are transformed as \r{dualb}, the dispersion equation \r{disp} is also invariant. 

Excluding zero and infinite values of the parameter $B_2$ and defining the transformation parameter $\VF$ to satisfy
\e \cot\VF=B_2, \f
the conditions \r{GBC1''} and \r{GBC2''} can be expressed as
\e \#a_1\.\#E_d = 0, \ \ \ \ \#a_2\.\#E_d = 0, \l{b2E}\f
or,
\e \#m\x\#E_d=0. \f
These conditions can be recognized as those of the E-boundary \cite{AP17,GBC}, which is a generalization of the PEC boundary. On the other hand, if we define the parameter $\VF$ by
\e \tan\VF = -B_2, \f
the conditions \r{GBC1''} and \r{GBC2''} can be expressed as
\e \#a_1\.\#H_d = 0, \ \ \ \ \#a_2\.\#H_d = 0, \f
or,
\e \#m\x\#H_d=0, \f
which correspond to those of the H-boundary \cite{AP17,GBC}, which is a generalization of the PMC boundary.  

In conclusion, the dispersion equation \r{disp} is reduced to one of the first order in $\#k$ when the boundary belongs to the class of generalized perfect electromagnetic conductor (GPEMC) boundaries, defined conditions of the form \r{GBC1''} and \r{GBC2''} or the form \r{GPEMC}. In this case, the possible $\#k$ vectors of a matched wave satisfy \r{mk0}. The GPEMC boundary can be interpreted as a duality-transformed E-boundary or H-boundary.

\subsection{Special Case}

Let us consider the special GPEMC boundary defined by a real unit vector $\#m$. Assuming a complex wave vector with real and imaginary parts,
\e \#k=\#k_{re} + j\#k_{im}, \f
for matched waves satisfying the dispersion condition \r{mk0}, both $\#k_{re}$ and $\#k_{im}$ must lie in the plane orthogonal to $\#m$, which is different from the plane of the boundary, in general. From $\#k\.\#k=k_o^2$ we obtain
\e \#k_{re}\.\#k_{re}-\#k_{im}\.\#k_{im}= k_o^2, \f
\e \#k_{re}\.\#k_{im}=0. \f
Assuming the $x,y,z$ coordinate system with $\#m=\#u_x$ we can assume $k_y$ known in $\#k=\#u_y k_y+ \#u_z k_z$, whence $k_z$ is obtained from
\e k_z = \sqrt{k_o^2- k_y^2}. \f
This is visualized by Figure \ref{fig:wavevectors}.
\begin{figure}
	\centering
	\includegraphics[width=70mm]{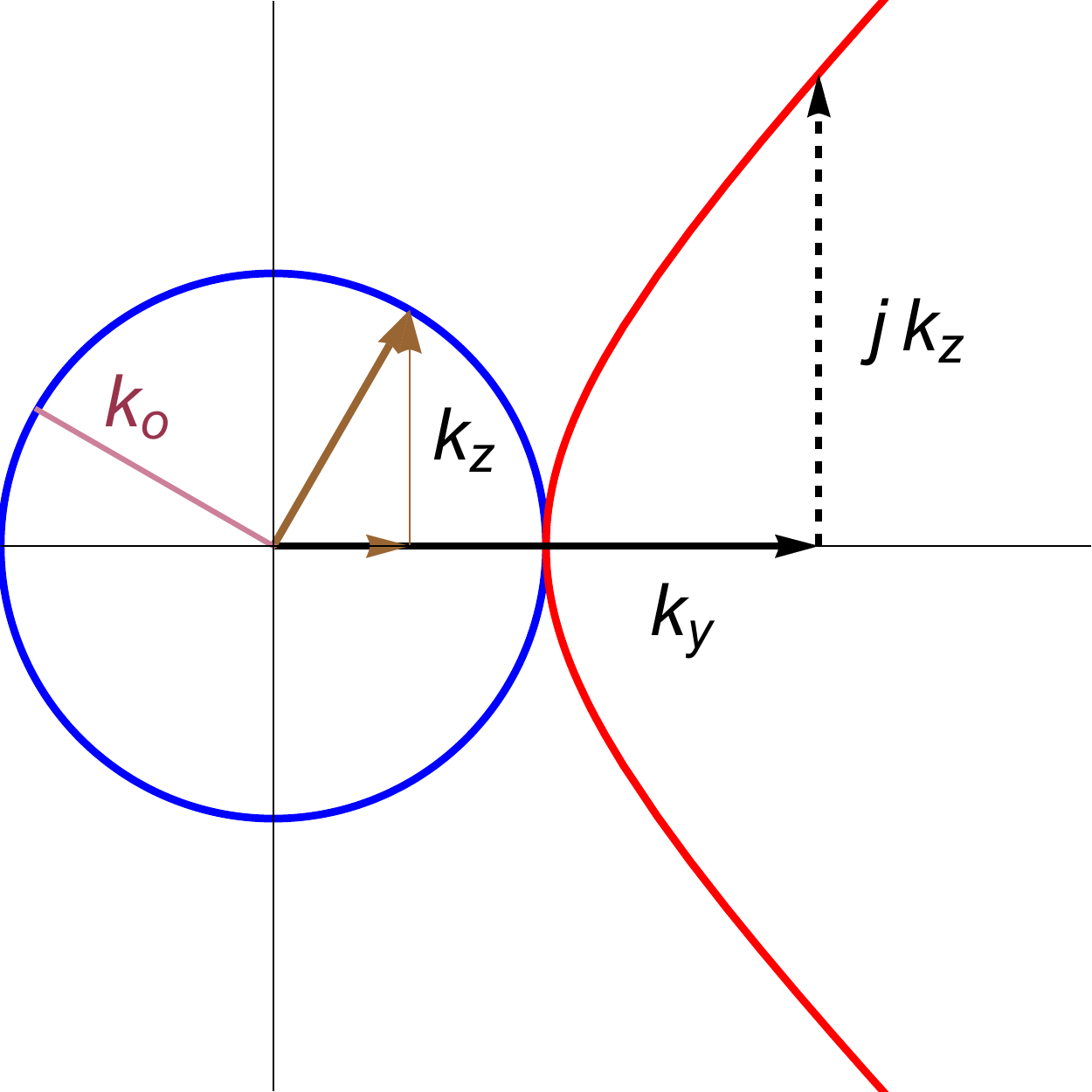}
	\caption{Plot of the wave vector $\#k$ associated to a matched wave for a GPEMC boundary is visualized in terms of a circle and a hyperbola. The vector $\#m=\#u_x$ is real and pointing towards the spectator. For real $k_y$, $\#k$ is real for $k_y<k_o$ and complex for $k_y>k_o$.}
	\label{fig:wavevectors}
\end{figure}

\subsection{Reflection from GPEMC Boundary}

Assuming an incident wave with the electric field
\e \#E^i(\#r) = \#E^i \exp(-j\#k^i\.\#r), \f
for a GPEMC boundary defined by \r{GPEMC} with a real vector $\#m$,
\e \#E^r(\#r) = \#E^r \exp(-j\#k^r\.\#r), \f
the reflected field can be found through the reflection dyadic $\=R$ as
\e \#E^r = \=R\.\#E^i. \f
The expression of the reflection dyadic can be written as (See Appendix 2),
\e \=R = \frac{-1}{(1+M^2\h_o^2)k_o^2\#m\.\#k^r}\#k^r\x\=K{}^r\.(\#m\x\=I)\.\=K{}^i, \l{R}\f
with
\e \=K{}^r = \#k^r\x\=I- k_oM\h_o\=I,\ \ \ \ \=K{}^i = \#k^i\x\=I+ k_oM\h_o\=I. \f

As a check of \r{R}, let us assume $|M|\ra\oo$, which corresponds to the special case of E-boundary \cite{GBC}. Expanding \r{R} yields
\e \=R= \frac{1}{\#m\.\#k^r}\#k^r\x(\#m\x\=I) = -\=I + \frac{\#m\#k^r}{\#m\.\#k^r}, \l{RE}\f
which coincides with Equation (5.245) of \cite{GBC}. The total field satisfies the condition
\e \#m\x(\#E^i+\#E^r)=\#m\x(\=I+\=R)\.\#E^i=0. \f
For $\#m=\#n$ the E-boundary is reduced to the PEC boundary.

Applying Equation (5.66) from \cite{GBC}, we can write for the reflected magnetic field component the rule
\e \#H^r = \frac{1}{k_o^2}(\#k^r\#k^i\xx\=R)\.\#H^i. \f
As another check, let us consider the case $M\ra0$. Substituting \r{R}, after some algebraic steps, we obtain
\e \#H^r \ra \frac{1}{\#m\.\#k^r}\#m(\#k^r\.\#H^i) - \#H^i, \f
whence the total field satisfies the condition of the H-boundary,
\e \#m\x(\#H^i+\#H^r)=0. \f
For $\#m=\#n$, this reduces to the condition of the PMC boundary.

\subsection{Polarization of Matched Wave}

The $\#k$ vector of a wave matched to a GPEMC boundary is any solution of \r{mk0}, $\#m\.\#k=0$. Any incident plane wave with zero reflection is matched. The field $\#E^i$ of a matched wave corresponding to a solution $\#k^i$ of \r{mk0} satisfies
\e \#k^r\x(\=K{}^r\.(\#m\x\=I)\.\=K{}^i)\.\#E^i=0. \f
Applying the dyadic rule \cite{Methods}
\e \=K{}^i\.\=K{}^{i(2)T} = (\det\=K{}^i)\=I, \l{det}\f
where the double-cross square and the determinant of the dyadic $\=K{}^i$ can be expanded as
\e \=K{}^{i(2)} = \#k^i\#k^i + k_oM\h_o\#k^i\x\=I + k_o^2M^2\h_o^2\=I, \f
and
\e \det\=K{}^i = k_o^3M\h_o(1+ M^2\h_o^2), \f
the polarization for the field of a matched wave can be expressed as
\ea \#E^i &=& E^i \=K{}^{i(2)T}\.\#m \\
&=& E^ik_oM\h_o( \#m\x\#k^i + k_oM\h_o\#m). \fa
To check this, because of \r{det}, we can expand
$$ \#k^r\x\=K{}^r\.(\#m\x\=I)\.\=K{}^i\.(\=K{}^{i(2)T}\.\#m) = $$
\e =\det\=K{}^i\#k^r\x\=K{}^r\.(\#m\x\#m)=0, \f
whence the field satisfies $\=R\.\#E^i=0$ and there is no reflected wave.

\subsection{Normal Incidence}

For a plane wave with normal incidence,
\e \#k^r = -\#k^i = k_o\#n, \f
we can substitute
\e \=K{}^r = -\=K{}^i = k_o(\#n\x\=I- M\h_o\=I),\f
in the expression of the reflection dyadic \r{R}, which is now reduced to
\e \=R = \frac{\#n}{A\#m\.\#n}\x\le(\#n\x\=I- M\h_o\=I)\.(\#m\x\=I)\.(\#n\x\=I- M\h_o\=I)\ri, \l{R2}\f
with $A=1+M^2\h_o^2$. Expanding this and noting that $\#n\.\#E^i=0$, we obtain the relation
\ea \#E^r &=&  \frac{1-M^2\h_o^2}{1+M^2\h_o^2}\#E^i  +  \frac{2M\h_o}{1+M^2\h_o^2}\#n\x\#E^i \\ 
&=&  \frac{1-M^2\h_o^2}{1+M^2\h_o^2}\#E^i  -  \frac{2M\h_o}{1+M^2\h_o^2}\h_o\#H^i . \l{ErHi}\fa
It appears remarkable that the GPEMC vector $\#m$, real or complex, does not play any role in normal incidence. Actually, \r{ErHi} reproduces the reflection rule for the PEMC boundary with $\#m=\#n$ (\cite{GBC}, Equation (2.36)).

\subsection{Numerical Examples}

As an example, let us consider a GPEMC boundary defined by $\#m=\#u_x\sin(\pi/3)+\#u_z\cos(\pi/3)$. The incident wave has unit amplitude and varying angle of incidence, $\#k^i/k_o= \#u_x\sin\theta-\#u_z\cos\theta$. Figure~\ref{fig:GPEMC1} illustrates the magnitude of the reflected wave for different polarizations. For $M=0$, the matched-wave condition can be seen to occur for the linear (perpendicular) polarization when $\#m\cdot\#k^i=0$. However, for $M\eta_o=1$, the polarization of the matched wave is no longer linear, and the two reflection coefficients are equally strong for all incidences. 

As another example, the GPEMC surface is defined by randomly generated complex $\#a_1$ and $\#a_2$ vectors, yielding an $\#m$ vector with complex components as
\ea \#m&=&(0.0682569 - 0.243121 j)\#u_x \nonumber\\
&+&(-0.397047+0.364515 j)\#u_y \nonumber\\
&+&(0.25906  +0.0128787j)\#u_z.\fa
Figure~\ref{fig:GPEMC2} displays the reflection characteristics when the angle of incidence is fixed as $(\theta=5\pi/12=75^\circ)$ and the azimuth angle $\varphi$ varies over the $2\pi$ range. The GPEMC parameter in this example is $M\eta_o=1.5$. There is no matched wave in this particular example.

\begin{figure}
	\centering
	\includegraphics[width=70mm]{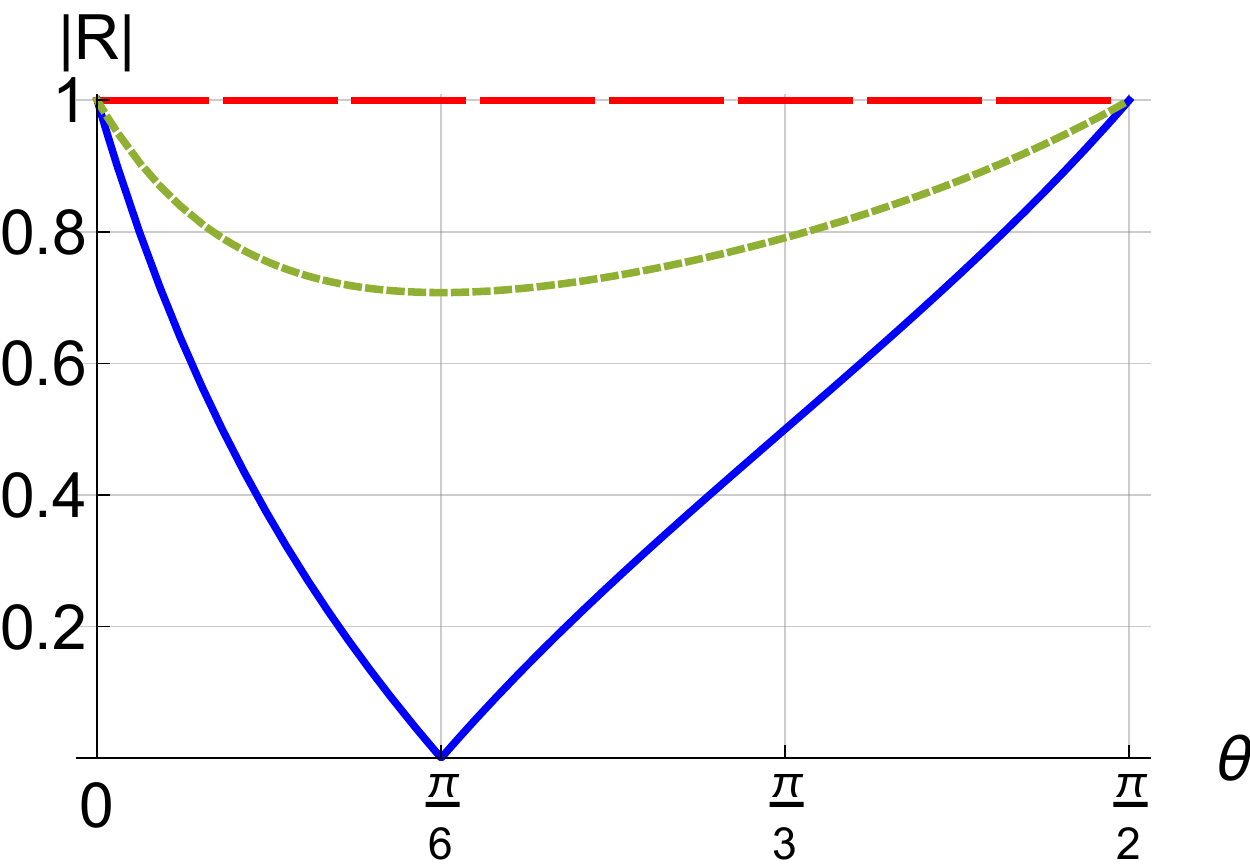}
	\caption{Magnitude of the reflection coefficient for perpendicular (solid blue) and parallel (dashed red) polarized plane wave, reflecting from a GPEMC surface with $M=0$ for varying angle of incidence, $\TH$. Dotted green line shows the reflection magnitude for $M\eta_o=1$ (same for both polarizations). The incident wave vector $\#k^i$ is along $\#u_x\sin\theta-\#u_z\cos\theta$ and the GPEMC vector $\#m=\#u_x\sin(\pi/3)+\#u_z\cos(\pi/3)$. Zero reflection corresponds to wave matched for the angle of incidence $\TH=\pi/6$.}
	\label{fig:GPEMC1}
\end{figure}

\begin{figure}
	\centering
	\includegraphics[width=70mm]{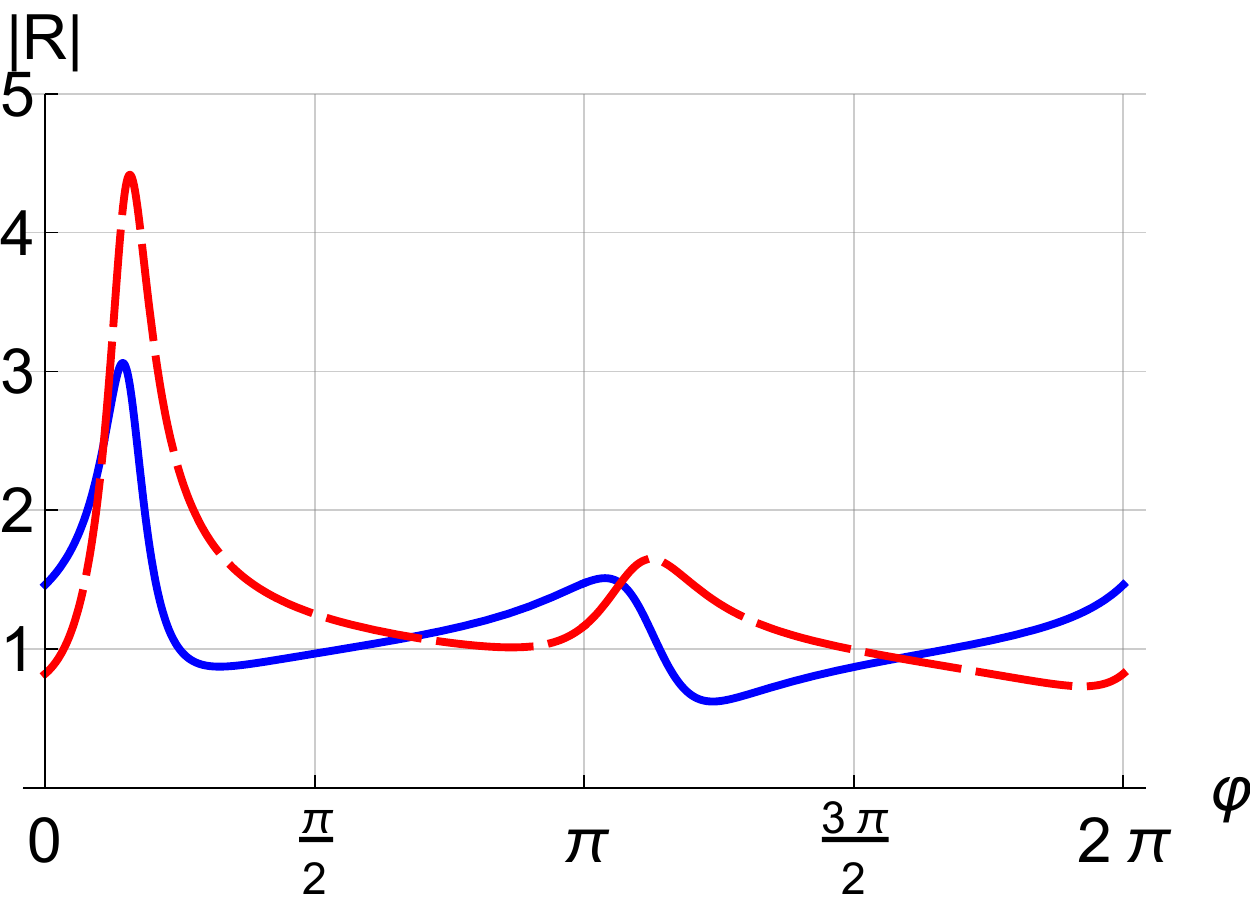}
	\caption{Magnitude of the reflection coefficient for perpendicular (solid blue) and parallel (dashed red) polarized wave for a GPEMC surface with $M\eta_o=1.5$ as function of the azimuth angle $\varphi$, with a fixed angle of incidence, $\theta=5\pi/12=75^\circ$. The GPEMC vector $\#m$ is a randomly generated complex vector.}
	\label{fig:GPEMC2}
\end{figure}

\section{No Dispersion Equation (NDE)}

Let us finally consider the problem of defining conditions for the GBC boundary allowing matched waves for any vector $\#k=k_o\#u$. Because \r{disp} is now an identity, let us call such a boundary as NDE boundary. An example was found in the previous Section as two special cases of the GPEMC boundary for $M =B_2/\h_o=\pm j/\h_o$. Electromagnetic media with no dispersion equation have been labeled in the past as NDE media (\cite{Multiforms}, Chapter 10).

To find other possible solutions, let us start by requiring that the dispersion equation \r{disp} written as
\e (\#a_1\#b_2-\#b_1\#a_2):(\#u\#u-\=I) + (\#a_1\x\#a_2  + \#b_1\x\#b_2)\.\#u =0, \l{dispu}\f
be valid for any unit vector $\#u$. Changing the sign of $\#u$, the sign of the last term of \r{dispu} is changed, whence the condition can be split in two parts as
\ea (\#a_1\#b_2-\#b_1\#a_2):(\#u\#u-\=I) &=& 0, \l{dispu1}\\
 (\#a_1\x\#a_2  + \#b_1\x\#b_2)\.\#u&=&0, \l{dispu2}\fa
each of which must be valid for any unit vector $\#u$. Obviously, \r{dispu2} requires
\e \#a_1\x\#a_2  + \#b_1\x\#b_2=0. \l{a12b12}\f
Choosing consecutively $\#u=\#u_1,\#u_2,\#u_3$ as three vectors making an orthonormal basis, summing the corresponding three conditions \r{dispu1} with $\sum (\#u_i\#u_i-\=I)=-2\=I$, yields 
\e (\#a_1\#b_2-\#b_1\#a_2):\=I=0, \f
whence \r{dispu1} requires
\e (\#a_1\#b_2-\#b_1\#a_2):\#u\#u=0 \f
for any $\#u$. From reasons similar to those of the previous Section, the symmetric part of the dyadic $\#a_1\#b_2-\#b_1\#a_2$ must be zero, whence the previously obtained condition \r{cond2} must be valid. Thus, the relations of the form \r{2} must be valid between the four vectors defining the NDE boundary.

Substituting \r{2} to the condition \r{a12b12}, we arrive at 
\e \#a_1\x\#a_2+ \#b_1\x\#b_2 = \le 1 + \frac{1}{B_2^2}\ri\#a_1\x\#a_2=0. \f
The case $\#a_1\x\#a_2=0$ applied to \r{GBC1''} and \r{GBC2''} would lead to an incomplete set of boundary conditions. Thus, the NDE boundary requires 
\e \ B_2 = \pm j. \f
In conclusion, boundary conditions for which matched waves satisfy the dispersion equation for any $\#k=k_o\#u$ must be of the form
\e \#m\x(\#H \pm (j/\h_o)\#E)=0, \l{NDE} \f
whence, there are no solutions beyond the two special cases $M=\pm j/\h_o$ of the GPEMC boundary.

Let us check this result. Assuming boundary conditions of either of the two forms in \r{NDE} and inserting
\e \#b_1 = \#a_1/B_2=\mp j\#a_1,\ \ \ \#b_2=\#a_2/B_2= \mp j\#a_2, \f
in the dispersion equation \r{D}, and expanding  
\e (k_o\#a_1 \mp j\#a_1\x\#k)\x(k_o\#a_2 \mp j\#a_2\x\#k)\.\#k=0  \f
term by term, it can be identified as being an identity. 

The case, $\#m=\#n$ of \r{NDE}, corresponding to two special cases of the PEMC boundary, was previously noticed in \cite{GBC} to define a boundary with no dispersion equation.

\section{Conclusion}

The dispersion equation governing matched plane waves associated to boundaries obeying general boundary conditions (GBC) has been studied for its special cases. In general, the dispersion equation is of the second order in the wave vector $\#k=k_o\#u$, defined by the unit vector $\#u$. Restrictions to the boundary conditions in the case when the dispersion equation is reduced to one of the first order were studied, and the boundaries were found to define a novel class for which the name generalized perfect electromagnetic conductor (GPEMC) was suggested. The $GPEMC$ boundary is defined by a vector $\#m$ with arbitrary magnitude. When $\#m$ is real and normal to the boundary, GPEMC equals the previously studied PEMC boundary.  An expression for the reflection dyadic corresponding to plane-wave reflection from the GPEMC boundary was derived and a few numerical examples were considered. For normal incidence, the GPEMC boundary turns out to act as the PEMC boundary for any vector $\#m$. Finally, boundary conditions for which there is no dispersion equation (NDE) (because it is identically satisfied by any $\#k$), were studied to define the class of NDE boundaries. It was found to be a certain special case of the class of GPEMC boundaries.

\section{Appendix 1: Duality Transformation}

In its basic form, duality transformation, based on the symmetry of the Maxwell equations, swaps electric and magnetic quantities. More generally, it is based on the linear transformation \cite{Methods}
\e \am \#E_d\\ \h_o\#H_d\a = \amm A & B \\ C & D\a \am \#E\\ \h_o\#H\a, \l{dual}\f
with $AD-BC\not=0$. The transformation changes fields, sources and conditions of electromagnetic media and boundaries. Choosing 
\e A=D=\cos\VF,\ \ \ B=-C=\sin\VF, \l{ADBC} \f
where $\VF$ is the transformation parameter, the simple isotropic medium is invariant \cite{GBC}, while the vectors defining the GBC boundary conditions \r{GBC1}, \r{GBC2} are transformed as
$$ \amm \#a_{1d} & \#b_{1d}\\ \#a_{2d} & \#b_{2d}\a = \amm \#a_1 & \#b_1\\ \#a_2 & \#b_2\a \amm D & -B \\ -C & A \a$$
\e = \amm \cos\VF\ \#a_1+ \sin\VF\ \#b_1  & -\sin\VF\ \#a_1+ \cos\VF\ \#b_1 \\ \cos\VF\ \#a_2+ \sin\VF\ \#b_2 & -\sin\VF\ \#a_2+ \cos\VF\ \#b_2\a. \l{dualb}\f
Applying this, one can find the relations
\ea \#a_{1d}\#b_{2d} - \#b_{1d}\#a_{2d} &=& \#a_1\#b_2- \#b_1\#a_2, \\
\#a_{1d}\x\#a_{2d} + \#b_{1d}\x\#b_{2d} &=& \#a_1\x\#a_2 + \#b_1\x\#b_2, \\
\#a_{1d}\.\#b_{2d} - \#b_{1d}\.\#a_{2d} &=& \#a_1\.\#b_2- \#b_1\.\#a_2,\fa
whence the dispersion equation \r{disp} is invariant in the duality transformation, $\#k_d=\#k$. Thus, the wave vector of a matched wave does not change in the duality transformation \r{dual}, \r{ADBC} of the boundary conditions.

\section{Appendix 2: Reflection Dyadic for GPEMC Boundary}

The reflection dyadic for the GPEMC boundary can be recovered from that of the more general GBC boundary by applying the expression from \cite{GBC}, eq. (5.63),
\e \=R = \frac{1}{J^r}\#k^r\x\=T. \l{R1}\f
Here we denote
\ea J^r&=& \#c_1^r\x\#c_2^r\.\#k^r, \\
\=T &=& \#c_2^r\#c_1^i-\#c_1^r\#c_2^i. \fa
The vector functions are defined by
\ea \#c_j^r &=&  \#k^r\x\#b_j - k_o\#a_j \\
\#c_j^i &=&  \#k^i\x\#b_j - k_o\#a_j.\fa
Substituting  $\#b_j=\#a_j/M\h_o$ for $j=1,2$, they become
 \ea \#c_j^r &=& \frac{1}{M\h_o}(\#k^r\x\=I-k_oM\h_o\=I)\.\#a_j, \\
\#c_j^i &=& \frac{1}{M\h_o}(\#k^i\x\=I-k_oM\h_o\=I)\.\#a_j, \fa

Applying \r{GBC1''} -- \r{GPEMC}, we can expand after some algebraic steps,
\e J^r = \frac{1+M^2\h_o^2}{M^2\h_o^2}k_o^2\#m\.\#k^r, \f
and
\e \=T=
\frac{-1}{M^2\h_o^2}(\#k^r\x\=I-k_oM\h_o\=I)\.(\#m\x\=I)\.(\#k^i\x\=I+ k_oM\h_o\=I). \f


\begin{thebibliography}{99}

\bibitem{PIER} I.V. Lindell and A. Sihvola, "Electromagnetic boundaries with PEC/PMC equivalence", {\it Prog. Electromag. Res. Lett.}, Vol. 61, pp. 119--123, 2016.

\bibitem{AP17} I.V. Lindell and A. Sihvola, "Electromagnetic wave reflection from boundaries defined by general linear and local conditions," {\it IEEE Trans. Antennas Propagat.}, Vol. 65, No. 9, pp. 4656 -- 4663, Sept. 2017.

\bibitem{GBC} I.V. Lindell and A. Sihvola, {\it Boundary Conditions in Electromagnetics}, Hoboken N.J.: Wiley and IEEE Press, 2020.

\bibitem{Methods} I.V. Lindell, {\it Methods for Electromagnetic Field Analysis}, 2nd ed., Oxford: University Press, 1995.

\bibitem{Hoppe} D.J. Hoppe and Y. Rahmat-Samii, {\it Impedance Boundary Conditions in Electromagnetics}, Washington, D.C.: Taylor and Francis, 1995.

\bibitem{Kildal} P.-S. Kildal, "Definition of artificially soft and hard surfaces for electromagnetic waves", {\it Electron. Lett.}, Vol. 24, pp. 168--170, 1988.

\bibitem{SH} P.-S. Kildal, "Artificially soft and hard surfaces in electromagnetics", {\it IEEE Trans. Antennas Propagat.}, Vol. 38, No. 10, pp. 1537--1544, Oct. 1990.

\bibitem{Rumsey} V.H. Rumsey, "Some new forms of Huygens' principle", {\it IRE Trans. Antennas Propag.}, Vol. 7, pp. S103--S116, Dec. 1959.

\bibitem{DB} I.V. Lindell and A. Sihvola, "Electromagnetic DB boundary", {\it Proc. XXXI Finnish URSI Convention}, Espoo, October 2008, pp. 81--82.

\bibitem{259} I.V. Lindell and A. Sihvola, "Electromagnetic boundary conditions defined in terms of normal field components," {\it IEEE Trans.\ Antennas Propag.}, Vol.58, no.4, pp.1128--1135, April 2010. 


\bibitem{SHDB} I.V. Lindell and A. Sihvola, "Soft-and-hard/DB boundary conditions realized by a skewon-axion medium," {\it IEEE Trans Antennas Propag.}, Vol. 61, No. 2, pp. 768--774, Feb. 2013.


\bibitem{PEMC} I.V. Lindell and A. Sihvola, "Perfect electromagnetic conductor", {\it J. Electro. Waves Appl.}, Vol.19, No.7, pp.861--869, 2005.

\bibitem{253} A. Sihvola and I. V. Lindell, "Perfect electromagnetic conductor medium." {\it Ann. Phys.} (Berlin) Vol.17, pp.787--802, September/October 2008.

\bibitem{264} A. Sihvola and I.V. Lindell, "Bianisotropic materials and PEMC," Chapter 26 in {\it Metamaterials Handbook, Theory and Phenomena of Metamaterials}, Boca Raton: CRC Press, pp.26.1--26.7, 2009.

\bibitem{Caloz13}   C. Caloz, A. Shahvarpour, D. L Sounas, T. Kodera, B. Gurlek and N. Chamanara, "Practical realization of perfect electromagnetic conductor (PEMC) boundaries using ferrites, magnetless non-reciprocal metamaterials (MNMs) and graphene," {\it Proc. URSI EMTS}, pp. 652--655, Hiroshima May 2013. 

\bibitem{Elmaghrabi} H. M.  El-Maghrabi,  A. M. Attiya and E. A. Hashish, "Design of a perfect electromagnetic conductor (PEMC) boundary by using periodic patches," {\it Prog.\ Electromag.\ Res.\ M}, Vol.16, pp.159--169, 2011.

\bibitem{235} I.V. Lindell and A.H. Sihvola, "Realization of the PEMC boundary," {\it IEEE Trans.\ Antennas Propag.},  Vol.53, no.9, pp.3012-3018, September 2005.

\bibitem{Wallen} H. Wall\'en and A. Sihvola: "How well can a PEC-backed gyrotropic layer approximate the ideal PEMC boundary?", {\it Proc.\ EuCAP 2006}, November 6-10, 2006, Nice, France, paper 349675hw (6 pages).

\bibitem{257} I.V. Lindell and A. Sihvola, "Electromagnetic boundary condition and its realization with anisotropic metamaterial," {\it Phys.\ Rev.\ E}, Vol.79, no.2, 026604 (7 pages), 2009.

\bibitem{258} I.V. Lindell and A. Sihvola, "Uniaxial IB-medium interface and novel boundary conditions," {\it IEEE Trans.\ Antennas Propag.}, Vol.57, no.3, pp.694--700, March 2009.

\bibitem{Zaluski11} D. Zaluski, D. Muha and S. Hrabar, "DB boundary based on resonant metamaterial inclusions," Proc. {\it Metamaterials 2011}, Barcelona, October, pp. 820--822, 2011.

\bibitem{274} I.V. Lindell and A. Sihvola, "Simple skewon medium realization of DB boundary condition,"  {\it Prog.\ Electromag.\ Res. Letters}, Vol.30, pp.29--39, 2012.

\bibitem{279} I.V. Lindell and A. Sihvola, "SHDB Boundary Conditions Realized by Pseudochiral Media," {\it IEEE Antennas Wireless Propag.\ Lett.}, Vol.12, pp.591--594, 2013.

\bibitem{Zaluski14}D.  Zaluski, S. Hrabar and D. Muha, "Practical realization of DB metasurface," {\it Appl.\ Phys.\ Lett.}, Vol.\ 104, 234106, 2014.

\bibitem{Kildal09}  P.-S. Kildal, "Fundamental properties of canonical soft and hard surfaces, perfect magnetic conductors and the newly introduced DB surface and their relation to different practical applications including cloaking," {\it Proc.\ ICEAA'09}, Torino, Italy Aug. 2009, pp. 607--610.

\bibitem{234} I.V. Lindell and A.H. Sihvola, "Transformation method for problems involving perfect electromagnetic conductor (PEMC) structures," {\it IEEE Trans.\ Antennas Propag.}, Vol.53, no.9, pp.3005-3011, September 2005.

\bibitem{241} I.V. Lindell and A. Sihvola, "Electromagnetostatic image theory for the PEMC sphere,"  {\it IEE Proc.\ Sci.\ Meas.\ Tech.}, Vol.153, no.3, pp.120-124, May 2006.

\bibitem{244} I.V. Lindell and A.H. Sihvola, "The PEMC resonator,"  {\it J. Electro. Waves Appl.}, Vol.20, no.7, pp.849--859, 2006.

\bibitem{245} I.V. Lindell and A.H. Sihvola, "Losses in the PEMC boundary,"  {\it IEEE Trans.\ Antennas Propag.}, Vol.54, no.9, pp.2553--2558, September 2006.

\bibitem{251} I.V. Lindell and A.H. Sihvola, "Reflection and transmission of waves at the interface of perfect electromagnetic conductor (PEMC)," {\it PIER B}, Vol.5, pp.169--183, 2008.

\bibitem{Sihvola06} A. Sihvola and I.V. Lindell, "Perfect electromagnetic conductor as building block for complex materials," {\it Electromagnetics}, Vol. 26, Nos. 3-4, pp. 279--287, April--June 2006.

\bibitem{250} A. Sihvola, P. Yl\"a-Oijala and I.V. Lindell, "Scattering by perfect electromagnetic conductor (PEMC) spheres: surface integral equation approach," {\it ACES Journal}, Vol.22, no.2, pp.236--249, July 2007.

\bibitem{Sihvola07} A. Sihvola, P. Yl\"a-Oijala and I.V. Lindell, " Scattering by PEMC (Perfect Electromagnetic Conductor) spheres using surface integral equation approach", {\it ACES, Applied Computational Electromagnetics Society Journal}, Vol. 22, No. 2, pp. 236-249, July 2007.

\bibitem{260} A. Sihvola, H. Wall\'en, M. Taskinen, P. Yl\"a-Oijala, H. Kettunen and I.V. Lindell, "Scattering by DB spheres," {\it IEEE Antennas Wireless Propag.\ Lett.}, Vol.8, pp.542--545, June 24, 2009. 

\bibitem{280} I.V. Lindell and A. Sihvola, "Surface waves on SHDB boundary," {\it IEEE Antennas Wireless Propag. Lett.}, Vol.13, pp.1027--1030, 2014.

\bibitem{285} I.V. Lindell and A. Sihvola, "Generalized Soft-and-Hard/DB Boundary" {\it IEEE Trans. Antennas Propag.}, Vol. 65, No. 1, pp. 226 -- 233, January 2017. 

\bibitem{Nosrati} M. Nosrati,  Z. Abbasi, M. Baghelani, S.Bhadra and M. Daneshmand, "Locally Strong-Coupled Microwave Resonator Using PEMC Boundary for Distant Sensing Applications", {\it IEEE Trans. Micro. Theory Tech.}, Vol. 67, No. 10, Oct. 2019.

\bibitem{Multiforms} I.V. Lindell, {\it Multiforms, Dyadics, and Electromagnetic Media}, Piscataway, N.J.: Wiley and IEEE Press, 2015.




\end{thebibliography}
\end{document}